\documentclass[12pt]{article}  
\usepackage{english}
 \newcommand{\beq}{\begin{equation}}
 \newcommand{\eeq}{\end{equation}}
 \newcommand{\beqa}{\begin{eqnarray}}
 \newcommand{\eeqa}{\end{eqnarray}}

 \def\dabl#1#2{\frac{{\rm d}{#1}}{{\rm d}{#2}}}
 
\begin{document}
\begin{center}
{\Large Using path integrals to price interest rate derivatives}
\vspace{5cm}

Matthias Otto \\[12pt]
Institut f\"ur Theoretische Physik, Universit\"at G\"ottingen\\
Bunsenstr. 9, D-37073 G\"ottingen
\end{center}
\vspace{10ex}
\noindent
{\bf Abstract}:
We present a new approach for the pricing of interest rate derivatives 
which allows a direct computation of option premiums without deriving a (Black-Scholes type) 
partial differential equation and without explicitly solving the stochastic process for the 
underlying variable. The approach is tested by rederiving the prices of a 
zero bond and a zero bond option for a short rate environment which
is governed by Vasicek dynamics. Furthermore, a generalization of the method to
general short rate models is outlined. In the case, where analytical
solutions are not accessible, numerical implementations of the path
integral method in terms of lattice calculations as well as path
integral Monte Carlo simulations are possible.

\vspace{2ex}
\noindent

\vspace{2ex}
\noindent

\newpage
\setcounter{equation}{0}
\noindent
\section{Introduction}
The purpose of this article is to present a new approach for the pricing of interest
rate derivatives: the path integral formalism. The claim is that interest 
rate derivatives can be priced without explicitly solving for the stochastic 
process of the underlying (e.g. a short rate) and without deriving a (Black-Scholes 
type) partial differential equation (PDE).

The mathematical foundation of the PDE approach to derivatives pricing, which is used 
traditionally, is the Feynman-Kac lemma \cite{feynman:48,kac:49} 
connecting the solution of a certain type of parabolic PDE to expectation values
with respect to stochastic processes. Usually, the original pricing problem 
is given in terms of an expectation value. Then the Feynman-Kac lemma is evoked 
in order to use the corresponding PDE to find the solution to the 
pricing problem. In the martingale approach to options pricing (see e.g.\cite{musiela:97}), 
usually the stochastic process for the underlying is solved, and the option price 
is determined using 
stochastic calculus.

The path integral formalism widely used in theoretical physics and first 
introduced in this field by Feynman \cite{feynman:48} does not need any reference to a Black-Scholes type 
equation nor does it depend on an explicit solution of the 
stochastic process for the underlying. It is traditionally introduced, however, 
by showing that a certain path integral satisfies a particular PDE which is known
to describe the problem under investigation. This procedure is also followed by 
Linetsky \cite{linetsky:98} who reviewed the early application of path integrals to 
finance (initiated by Dash \cite{dash:88,dash:89}) and who discusses
various analytically tractable payoffs . In statistical physics, 
the application of path integrals to option pricing has been recently discussed by Baaquie 
who derived a Schr\"odinger equation for the corresponding pricing
problem \cite{baaquie:97}. Let us mention here, that topics in finance
have become increasingly
popular in theoretical physics (see e.g. Bouchaud
\cite{bouchaud:98} and references therein).

As opposed to the approach followed in \cite{linetsky:98,baaquie:97}, in this article, we introduce the path integral from a stochastic viewpoint only
(using the Martin-Siggia-Rose (MSR) formalism \cite{martin:72,graham:73,dedominicis:76,janssen:76}) 
without making any appeal to particular PDEs. In our view, this procedure 
is easier to carry out than proving that a specific path integral satisfies 
a particular PDE. The MSR method is well known in statistical physics
and finds a rigorous justification in the context of the
Cameron-Martin-Girsanov theorem from stochastic analysis \cite{oksendal:98}.

The path integral approach is introduced here in the context of interest rate
products. It has been discussed from a more general perspective by Linetsky \cite{linetsky:98}.
In order to be specific, the framework is presented for the short rate
model first introduced by Vasicek. 
The fact that analytical results exist for this model, permits a serious test 
of our approach. In particular, we value zero bonds and zero bond options as special 
cases of path-dependant contingent 
claims.

In principle, every short rate model can be cast into a path integral. 
More generally, any process describing an underlying random variable 
(tradable or non-tradable) that can be modelled in terms of a Langevin 
equation (or in mathematical terms, a stochastic differential
equation)  can be rewritten in this approach. 
The path integral might itself serve as a starting point to state new 
models. The criterion of no-arbitrage that has to be fulfilled by any 
model (also by those presented in a path integral from) is slightly touched on in 
section 7.

The path integral approach is not just a new way to obtain results already 
known, but it may serve as a general framework for numerical work. 
In fact, one of the key ingredients is a probability density functional 
for one or several stochastic underlyings. A European claim is valued by 
integrating over all possible paths taken by the underlying variables 
according to the weighting imposed by the functional. Moreover, a 
Monte Carlo simulation can be set up by sampling the different paths 
taken by the underlying according to the probability density functional.

The article is divided as follows. In section 2 we discuss the Vasicek model 
as a prototype model that can be stated in terms of a Langevin equation. 
Next, in section 3 the pricing of contingent claims is considered in a 
general context. Then in section 4 we derive a probability density functional 
for the short rate which we then, in section 5, apply to calculate 
the price of a simple contingent claim, a zero bond, and in section 6, to 
determine the price of a zero bond option. In section 7, we make a few remarks 
on the path integral as a new guise for (short rate) models, before 
reaching a conclusion.

\section{The Vasicek model as a Langevin equation}

The Vasicek model \cite{vasicek:77} in finance is usually studied in terms of a stochastic 
differential equation for the short rate $r$, i.e. the rate of instantaneous 
borrowing (and lending, assuming no bid-offer spreads):
\begin{equation}
\label{vas}
dr(t)=a(b-r(t))dt+\sigma dz(t)
\end{equation}
The parameters $a$ and $b$ model the mean reversion, 
$\sigma$ is the volatility of the short rate. Both $a$,$b$ and
$\sigma$ are constant in time. The increment $dz(t)$ represents a Wiener process. The whole dynamics
is parametrized in terms of the continuous time variable $t$.
Next, we write the equivalent Langevin equation \cite{karatzas:91}:
\begin{equation}
\label{langevin}
\dabl{r}{t}=a(b-r(t))+\sigma f(t)
\end{equation}
Crudely speaking, this equation can be obtained by dividing Eq.(\ref{vas}) by $dt$.
The stochastic "´force"´ $f(t)$ is a practical way to think of the 
mathematical Wiener increment $dz$ per unit time $dt$, so very roughly, 
we say that
\begin{equation}
\dabl{z}{t}\simeq f(t)
\end{equation}
The Langevin equation (\ref{langevin}) also describes the motion of
Brownian particle which is attached by a spring.

The fact  that $dz$ is a Wiener process implies that the force
$f(t)$ satisfies 
\begin{equation}
\label{white}
\langle f(t)f(t')\rangle=\delta (t-t'),\;\; \langle f(t)\rangle=0
\end{equation}
The brackets $\langle\dots\rangle$ indicate a mean value with respect to 
a Gaussian distribution of zero mean and variance of one.

\section{Contingent claims}
Using the terminology of \cite{rebonato:96},
the so-called classical Vasicek approach to pricing contingent claims 
is to derive a PDE subject to appropriate boundary conditions which 
model the payoff profile. The PDE is obtained using Ito's lemma and setting 
up a riskless portfolio which eliminates the Wiener process. The solution 
of the PDE then gives the price for the claim at time $t$ prior (and up) 
to maturity time $T$.

The Feynman-Kac lemma states that the solution of the particular PDE
considered in the classical approach is given as an expectation 
value with respect to the martingale measure obtained from the stochastic
process for the underlying tradable \cite{karatzas:91}. In our case this is 
the discounted bond price $Z(t,T)=B_t^{-1}P(t,T)$, where $P(t,T)$ is the (zero) bond 
price at time $t$ promissing to pay one unit at maturity $T$ and $B_t$ 
is a cash bond at time $t$ which has accrued form $B_0=1$ at time $0$. The martingale
measure for $Z(t,T)$ leads to a risk-adjusted process for the short rate
$r(t)$, which in the case of the Vasicek model leads merely to a renormalization 
of the parameter $ab$. Later on in the discussion, we will assume
that the parameters $a$ and $b$ are the risk-adjusted ones \cite{rebonato:96}.

Now by virtue of the Feynman-Kac lemma the price of any claim $V(t)$ at time 
$t$ whose payoff at time $T$ is $X$ is given by:
\beq
\label{expect}
V(t)=E_Q\left[e^{-\int_t^T ds r(s)}X|r(t)=r\right]
\eeq
The subscript $Q$ indicates the martingale measure.

Later on, we will consider the bond price $P(t,T)$ as an example of a claim 
with payoff $X=1$ at maturity $T$ and an option on a zero bond both of
which we will price using the path 
integral technique. Before, we set up the stage in order to be able to carry 
out the expectation value in Eq.(\ref{expect}).

\section{The distribution functional for the short rate}
We now provide the foundation for the path integral approach. In order 
to be specific, we will start from the stochastic process for the 
Vasicek model, or rather the corresponding Langevin equation Eq.(\ref{langevin}).
This equation is transformed using the so-called Martin-Siggia-Rose formalism 
\cite{martin:72,graham:73,dedominicis:76,janssen:76} to give a probability density functional for the short rate.

For a given realization of the force $f(t)$, we have the conditional 
probability density functional for all pathwise realizations of the short 
rate process given by Eq.(\ref{langevin}):
\begin{equation}
p\left(\{r(s)\}|\{f(s)\}\right)=
{\cal N}\prod_s
\delta\left(
\dabl{r}{s}-a(b-r(s))-\sigma f(s)
\right)
\end{equation}
The constant ${\cal N}$ is a normalization factor which is left unspecified.
Of course, we later constrain the parameter $s$ to range from $t$ to $T$, as well as 
the product on the r.h.s of the equation. In what follows, this
restriction $s\in[t,T]$ is understood.
Next, we sum over the "force" $f(t)$:
\begin{equation}
p\left(\{r(s)\}\right)=
{\cal N}\int{\cal D}f(s) e^{-\frac{1}{2}\int ds f^2 (s)}\prod_s
\delta\left(
\dabl{r}{s}-a(b-r(s))-\sigma f(s)
\right)
\end{equation}
The integration is weighted by a Gaussian as we imposed white noise correlations 
(see Eq.(\ref{white})). Obviously, the measure ${\cal D}f(s)$ is a
very peculiar one which we will comment on in more detail at the end
of the next 
section. For the moment, let us consider it as a sum within the space
of continuous real-valued functions $f(s)$.

Next we 
carry out the functional integral with respect to $f(s)$. For this
purpose, the delta function needs 
to be rewritten for fixed $s$ as follows:
\begin{equation}
\delta\left(
\dabl{r}{s}-a(b-r(s))-\sigma f(s)
\right)=
\frac{1}{\sigma}\delta\left(
\frac{1}{\sigma}\left(\dabl{r}{s}-a(b-r(s))\right)-f(s)
\right)
\end{equation}
The factor $1/\sigma$ can be absorbed 
into the
normalization constant because it is constant with respect to the
parameter $s$. For time-dependant volatilities $\sigma=\sigma(s)$,
which arise in general short rate models, the
situation is more complicated (see section 8). 

Finally, one obtains for the probability density functional for the short rate
$r(s)$:
\begin{equation}
\label{pdf}
p\left(\{r(s)\}\right)=
{\cal N}
e^{-\frac{1}{2\sigma^2}\int ds 
\left(\dabl{r(s)}{s}-a(b-r(s))\right)^2
}
\end{equation}
As mentioned, 
only the family of short rates within a specific time window is considered, 
e.g. between $t$ and $T$. Therefore, the integration on the r.h.s. of Eq.(\ref{pdf})
has to be constrained to the interval $[t,T]$. It is interesting to 
note here that negative short rates have a non-vanishing probability, 
a fact that is well-known for the Vasicek model. Alternative short rate
models exist which bypass this drawback. In fact, in the path integral
approach one could also enforce non-negative short rates by a delta
function which is a subject of future work.
The probability density functional can now be used to carry out the expectation 
value for any claim in Eq.(\ref{expect}).

\section{The bond price as a path integral}
Let us discuss the bond price as a specific example for a claim. 
The bond price today, i.e. at time $t$, for a monetary unit
 promissed at some later time $T$ can be stated as an expectation 
value of the 
discount function:
\begin{equation}
\label{bond}
P(t,T)=E_Q\left[e^{-\int_t^T ds r(s)}|r(t)=r\right]
\end{equation}
The classical Vasicek approach (as described in \cite{rebonato:96}) 
has been to write down the corresponding PDE for $P(t,T)$ rather than 
evaluating the expectation value in Eq.(\ref{bond}) directly.

Using the probability density functional in Eq.(\ref{pdf}), we are now in the position to evaluate the expectation value 
in Eq.(\ref{bond}) for $P(t,T)$. It is explicitly given as follows:
\begin{equation}
\label{P1}
P(t,T)=\frac{\int_{-\infty}^\infty dr(T)
\int_{r(t)}^{r(T)}{\cal D}r(s)
\exp\left(-\frac{1}{2\sigma^2}\int_t^T ds 
\left(\dabl{r(s)}{s}-a(b-r(s))\right)^2
-\int_t^T ds r(s)
\right)}
{\int_{-\infty}^\infty dr(T)
\int_{r(t)}^{r(T)}{\cal D}r(s)
\exp\left(-\frac{1}{2\sigma^2}\int_t^T ds 
\left(\dabl{r(s)}{s}-a(b-r(s))\right)^2
\right)}
\end{equation}
The argument of the expectation value Eq.(\ref{bond}) appears in the
numerator of the equation above as the
second term inside the exponential function. The first term stems from
the probability density functional Eq.(\ref{pdf}). The denominator
gives the proper normalization. The functional integrations in the
numerator are due to the fact that a conditional expectation value on
$r(t)=r$ is desired. Therefore, first a transition probability (modulo
a constant) for the short rate to pass form $r(t)$ to $r(T)$ is
calculated, and then an integration with respect to the final rate
$r(T)$ is carried out.
 
Let us now substitute for $x(s)=b-r(s)$. Then after carrying out
integrations involving boundary terms, Eq.(\ref{P1}) reads 
as
\begin{equation}
\label{bond2}
P(t,T)=\frac{X}{Y}
\end{equation}
with 
\begin{eqnarray}
X&=&\int_{-\infty}^\infty dx(T)\int_{x(t)}^{x(T)} {\cal D}x(s)
\exp\left(
-\frac{1}{2\sigma^2}\int_t^T ds 
\left(
\left(\dabl{x(s)}{s}\right)^2+a^2x^2(s)\right)
\right.\nonumber\\
&-&\left.\frac{a}{2\sigma^2}
(x^2(T)-x^2(t))
-b(T-t)+\int_t^T ds x(s)
\right)
\end{eqnarray}
and 
\begin{eqnarray}
Y&=&\int_{-\infty}^\infty dx(T)\int_{x(t)}^{x(T)} {\cal D}x(s)
\exp\left(
-\frac{1}{2\sigma^2}\int_t^T ds 
\left(
\left(\dabl{x(s)}{s}\right)^2+a^2x^2(s)\right)
\right.\nonumber\\
&-&\left.\frac{a}{2\sigma^2}
(x^2(T)-x^2(t))
\right)
\end{eqnarray}
The functional integrals over $x(s)$ can be evaluated 
using a familiar formula known in physics for the 
propagator of the harmonic oscillator
(quoted in the appendix). From this formula one obtains for the numerator:
\begin{eqnarray}
\label{num}
X&=&\sqrt{\frac{a}{2\pi \sigma^2 \sinh(a(T-t)}}\int_{-\infty}^\infty dx(T)
\exp\left(
\frac{\sigma^2}{2a^3}\left(
e^{-a(T-t)}-1+a(T-t)
\right)
\right.\nonumber\\
&-&\left.\frac{a}{2\sigma^2 \sinh(a(T-t))}
\left(
(x^2(T)+x^2(t))\cosh(a(T-t))
-2x(T)x(t)
\right.\right.\nonumber\\
&+&\left.\left.
2\left(e^{a(T-t)}-1\right)(C(x(T)+x(t))+C^2)
\right)
\right.\nonumber\\
&-&\left.\frac{a}{2\sigma^2}
(x^2(T)-x^2(t))
-b(T-t)
\right)
\end{eqnarray}
where 
\beq
C=\frac{i\sigma^2}{2a}\int_0^{T-t}du e^{-au}i=\frac{\sigma^2}{2a^2}
\left(e^{-a(T-t)}-1\right)
\eeq
Likewise one obtains an expression for the denominator:
\begin{eqnarray}
\label{denom}
Y&=&\sqrt{\frac{a}{2\pi \sigma^2 \sinh(a(T-t)}}
\int_{-\infty}^\infty dx(T)
\exp\left(-
\frac{a}{2\sigma^2}
(x^2(T)-x^2(t))
\right.\nonumber\\
&-&\left.\frac{a}{2\sigma^2 \sinh(a(T-t))}
\left(
(x^2(T)+x^2(t))\cosh(a(T-t))
-2x(T)x(t)\right)
\right)\nonumber\\
\end{eqnarray}
It remains to perform the (Gaussian) integration with respect to $x(T)$. 
In agreement with standard notation (see e.g. Hull \cite{hull:97}), we define 
\beq
B(t,T)=\frac{1-e^{-a(T-t)}}{a}
\eeq
The relation to our function $C$ is as follows:
\beq
C=-\frac{\sigma^2}{2a}B(t,T)
\eeq
Collecting terms from the Gaussian integration in Eq.s (\ref{num}) and 
(\ref{denom}) and substituting the results for $X$ and $Y$ in 
Eq.(\ref{bond2}), one finally obtains:
\beq
\label{bondfinal}
P(t,T)=A(t,T)\exp\left(-B(t,T)r\right)
\eeq
with
\beq
A(t,T)=\exp\left(\left(-B(t,T)+T-t\right)\left(\frac{\sigma^2}{2a^2}-b\right)-
\frac{\sigma^2}{4a}B^2(t,T)\right)
\eeq
This result is in full agreement with an earlier result by Vasicek who 
uses the PDE method (see \cite{rebonato:96}). As we have successfully tested 
the method for a known result, it is justified to use it for further applications.

Concerning the functional integrations performed above, the question
of 
existence
of the functional integration measure ${\cal D}r(s)$ and ${\cal
  D}f(s)$ needs to be addressed. A full treatment of this issue is
beyond the scope of this article (see e.g. \cite{oksendal:98}). Instead, we present a heuristic
approach. Let us consider a functional integration
$\int {\cal
  D}f(s)$
within  the time interval $[t,T]$ which corresponds to fixing the
functions $f(s)$ at times $t$ and $T$ to $f(t)$ and $f(T)$
respectively. We slice  the time interval $[t,T]$ in $N$ intervals of
lenght $\tau$ such that $T-t=N\tau$. Thus we obtain a sequence of
times $t_0=t<t_1<..<t_i<t_{i+1}<..<t_N=T$ with $i=0..N$. From a heuristic viewpoint, an integration operator
${\cal D}f(s)$ 
may be written as a limit:
\beq
\label{measure}
\int_{f(t)}^{f(T)}{\cal D}f(s)=\lim_{N\rightarrow\infty}\prod_{i=1}^{N-1}
\int df(t_i)
\eeq
The integrations on the r.h.s. are simple (Riemannian) 
integrations with
respect to real variables $f(t_i)$. In fact, the r.h.s. of
Eq.(\ref{measure}) is used for numerical evaluations of functional
integrals \cite{ryder:85}. In this sense, the sum over all paths taken
by the function $f(s)$ from $f(t)$ to $f(T)$ is approximated by a walk
with discrete steps in time.
Of course, one has to make sure that after performing the
limit $N\rightarrow\infty$ the results of functional integrations are independant of the
way the time interval was discretized. Moreover, one has to take 
care of diverging normalization factors (appearing e.g. in section
4). Since all results we are interested
in appear as expectation values and therefore as ratios of functional
integrals, possibly diverging normalizations cancel (which is easy to
show in a setting of
discrete time). A mathematically rigorous treatment of functional
integrals used for expectation values involves the
Cameron-Martin-Girsanov theorem \cite{oksendal:98}.

\section{The bond option price as a path integral}
As we propose the path integral approach in particular for the pricing 
of interest rate derivatives, let us demonstrate its usefulness for 
the example of bond options when short rates are again governed by Vasicek 
dynamics.

Specifically, let us consider the price of a European call at time $0$ with 
strike $k$ expiring at time $t$ on zero bond with maturity $T$, with $T>t$:
\beq
\label{call1}
c(0)=E_Q\left[
e^{-\int_0^t ds r(s)} \max\left(
P(t,T)-k;0
\right)
|r(0)=r
\right]
\eeq
The function $\max(x;0)$ gives the larger value of $x$ or 0. For 
Vasicek dynamics the European call price can be determined analytically. 
The r.h.s. of Eq.(\ref{call1}) vanishes unless $P(t,T)>k$. As in our 
case the zero bond price has the form
\beq
P(t,T)=A(t,T)e^{-B(t,T)r}
\eeq
this condition can be rewritten as:
\beq
r(t)<\frac{1}{B(t,T)}\ln\left(\frac{A(t,T)}{k}\right)=\tilde{k}
\eeq
The call price formula is then given by:
\beqa
\label{call2}
c(0)&=&A(t,T)E_Q\left[
e^{-\int_0^t ds r(s)-B(t,T)r}|r(t)<\tilde{k}\right]
\nonumber\\
&-&kE_Q\left[
e^{-\int_0^t ds r(s)}|r(t)<\tilde{k}\right]
\eeqa
We now follow the steps outlined in the previous section. First, 
we substitute $x(s)=b-r(s)$. The condition $r(t)<\tilde{k}$
 then translates into $x(t)>b-\tilde{k}$. The first expectation value 
on the r.h.s. of Eq.(\ref{call2}) is then given by:
\beq
E_Q\left[
e^{-\int_0^t ds r(s)-B(t,T)r}|r(t)<\tilde{k}\right]=\frac{X}{Y}
\eeq
with 
\beqa
\label{num2}
X&=&\int_{b-\tilde{k}}^\infty dx(t)\int_{x(0)}^{x(t)} {\cal D}x(s)
\exp\left(-
\frac{1}{2\sigma^2}\int_0^t ds\left(
\left(\dabl{x(s)}{s}\right)^2+a^2x^2(s)
\right)
\right.\nonumber\\
&-&\left. \frac{a}{2\sigma^2}
(x^2(t)-x^2(0))-b(T-t)+\int_0^t ds x(s)
-B(t,T)(b-x(t))
\right)\nonumber\\
\eeqa
and
\beqa
Y&=&\int_{-\infty}^\infty dx(t)\int_{x(0)}^{x(t)} {\cal D}x(s)
\exp\left(-
\frac{1}{2\sigma^2}\int_0^t ds\left(
\left(\dabl{x(s)}{s}\right)^2+a^2x^2(s)
\right)
\right.\nonumber\\
&-&\left. \frac{a}{2\sigma^2}
(x^2(t)-x^2(0))
\right)\nonumber\\
\eeqa
The condition $x(t)>b-\tilde{k}$ enters into numerator $X$ as a lower bound of integration 
with respect to $x(t)$. The next steps of integration are straightforward and follow along the lines 
of Eq.s(\ref{num}) to (\ref{bondfinal}). Useful relations are:
\beq
P(0,T)=E_Q\left[
e^{-\int_0^t ds r(s)} P(t,T)|r(0)=r\right]
\eeq
and 
\beq
P(0,t)=E_Q\left[
e^{-\int_0^t ds r(s)}|r(0)=r\right]
\eeq
which follow form the no-arbitrage condition Eq.(\ref{noarb}) to be discussed later. The bond prices
$P(0,T)$ and $P(0,t)$ make up for the two terms on the r.h.s. of Eq.(\ref{call2}) up to the factors 
$N(h)$ and $k N(h-\sigma_P)$ which result form the bounded integrations with respect to $x(t)$ 
(see Eq.(\ref{num2}). In fact, we rederive the Jamshidian's formula for a European call on a 
zero bond \cite{jamshidian:89}:
\beq
c(0)=P(0,T)N(h)-kP(0,t)N(h-\sigma_P)
\eeq
where $N(x)$ is a cumulative normal distribution and 
\beq
h=\frac{1}{\sigma_P}\ln\left(\frac{P(0,T)}{kP(0,t)}\right)+\frac{\sigma_P}{2}
\eeq
with
\beq
\sigma_P=\frac{\sigma}{a}\left(1-e^{-a(T-t)}\right)\sqrt{\frac{1-e^{-2at}}{2a}}
\eeq
For completeness we state the formula for a put which follows from put-call parity \cite{hull:97}:
\beq
p(0)=kP(0,t)N(-h+\sigma_P) - P(0,T) N(-h)
\eeq
The analytical tractability of the bond option formulas for Vasicek dynamics is of course dependant on 
the simple form of the equation for the bond price itself. In general, this simple form remains true for 
so-called {\it affine} models \cite{rebonato:96} if one abstracts from the specific expressions given for 
$A(t,T)$ and $B(t,T)$. It allows for a simple translation of the condition on the terminal bond price into a condition
on the terminal short rate. The latter can then be easily implemented as an integration boundary in the evaluation 
of the expectation value.

\section{Models for the short rate in a new guise}
Looking at the conditional probability density for the Vasicek model, 
Eq.(\ref{pdf}), we can take the negative argument of the exponential 
function as a starting point to state models for the short rate in 
a different form. Using an analogy to physics, we call it the "Hamiltonian"
 for the short rate which reads in the case of Vasicek's model as 
follows:
\beq
\label{ham.V}
H_{Vasicek}=\frac{1}{2\sigma^2}\int ds \left(
\dabl{r(s)}{s}-a(b-r(s))
\right)^2
\eeq
New models for the short rate can be given by stating a different function
$H$ or a different probability density functional. Of course, these new 
models must be arbitrage-free in complete markets in the sense of
\cite{harrison:81}. In order to do so \cite{rebonato:96}, they 
have to fulfil the following implicit constraint ($t\leq\tau\leq T$):
\beq
\label{noarb}
P(t,T)=E_Q\left[
e^{-\int_t^\tau ds r(s)}P(\tau,T)|r(t)=r
\right]
\eeq
The Hamiltonian enters implicitly into the constraint in calculation 
of $P(t,T)$ on the l.h.s. and in the calculation of $P(\tau,T)$ and the 
evaluation of the expectation value on the r.h.s. of the equation.
Eq.(\ref{noarb}) is known to hold (and can be verified explicitly using the 
path integral) for Vasicek dynamics.

It would be desirable to derive a simple criterion for the class 
of functions $H$ or the probability density functionals that are allowed by 
the no-arbitrage condition Eq.(\ref{noarb}). In particular, it would be 
challenging to consider probability density functionals which are not derived
from known short rate processes.

\section{The path integral for general short rate dynamics}
After having introduced the path integral approach for Vasicek
dynamics, we generalize the approach to any model for the short rate
which can be cast in the following form:
\beq
\label{general}
dr(t)=\rho(r(t),t)dt+\sigma(r(t),t)dz(t)
\eeq
After following the procedure of section 4, 
the equivalent ``Hamiltonian'' in the path integral framwork is
obtained:
\beq
\label{ham.gen}
H_{general}=\frac{1}{2}\int ds \frac{1}{\sigma(r(s),s)^2}
\left(\dabl{r(s)}{s}-\rho(r(s),s)
\right)^2
+\int ds \phi(s)\sigma(r(s),s)\phi^*(s)
\eeq
The first term in Eq.(\ref{ham.gen}) is an obvious generalization of 
Eq.(\ref{ham.V}), the Hamiltonian of the Vasicek model. The second
term, however, arises whenever $\sigma(r(s),s)$ is not constant, but
depends on the short rate level $r(s)$ and/or time $s$. The functions
$\phi(s)$ and $\phi^*(s)$ (the complex conjugate of $\phi(s)$) are random fields which need to be summed
over in addition to the functional integration with respect to the
short rate. In fact, every claim on a payoff $X$ at time $T$
is valued at time $t$ according to:
\beq
V(t)=E_Q\left[e^{-\int_t^T ds r(s)}X|r(t)=r\right]
\eeq
where the conditional expectation value is given by:
\beq
\label{pi.gen}
E_Q\left[Y|r(t)=r\right]=
\frac{\int_{-\infty}^\infty dr(T)
\int_{r(t)}^{r(T)}{\cal D}r(s)
\int {\cal D}\phi(s)\int {\cal D}\phi^*(s)
\exp\left(
-H_{general}
\right)Y}
{\int_{-\infty}^\infty dr(T)
\int_{r(t)}^{r(T)}{\cal D}r(s)
\int {\cal D}\phi(s)\int {\cal D}\phi^*(s)
\exp\left(-H_{general}
\right)}
\eeq
The integration boundaries for the random fields $\phi(s)$ and
$\phi^*(s)$ have been suppressed. In fact, in a discretized form of
the functional integral for a time interval $[t,T]$, the  fields $\phi_i$
and $\phi^*_i$ for $i=0..N$ for given points in time $t_i=t+i\tau$, where $N\tau=T-t$, are
defined. The integration with respect to the in general complex fields
$\phi_i$
and $\phi^*_i$ is mapped to an integration with respect to 
real and imaginary parts $Re(\phi_i)$, $Im(\phi_i)$  respectively,
with the integration boundaries ranging 
from $-\infty$ to $\infty$ \cite{ryder:85}. At the end of the
calculation the limit $N\rightarrow\infty$ and $\tau\rightarrow 0$ with
$N\tau=T-t$ finite is performed.

The functional integrations in Eq.(\ref{pi.gen}) can not be carried out
exactly, but need to be performed on a 
lattice using a
computer. However, this task can be accomplished in principle for any
short rate model specified by a particular set $\rho(r(t),t)$ and
$\sigma(r(t),t)$. This is the subject of future work.

\section{Conclusion}
We have presented a new approach to the pricing of interest rate
derivatives, 
based on the path integral well developed in theoretical physics.
It was tested for the simple case of a zero bond and a bond option in a
world where the short rate is governed by Vasicek dynamics. The familiar 
results of Vasicek and Jamshidian were recovered. 
The approach was generalized to arbitrary short rate models given as
a stochastic differential equation.
Other claims can be 
readily priced within the new framework. However, analytical solutions 
are not always available.

Complementing numerical methods for pricing derivatives 
which take PDE's as their starting point such as finite elements, the new approach can be put to 
use in numerical applications. One possible choice is the direct 
computation of the path integral on a lattice. Another application
would be in 
so-called path integral Monte Carlo simulations, which are 
also well established in physics \cite{binder:96}. Thus, the path
integral approach might prove to become a powerful tool derivatives
pricing and financial engineering.\\

\noindent
{\bf Acknowledgements:} The author acknowledges stimulating discussions
with C.T. Hille and
J. Topper from Arthur Andersen, Frankfurt,
where the first ideas to the present work emerged, O. D\"urst
from DG Bank, Frankfurt, and with colleagues at the Institut f\"ur
Theoretische Physik, G\"ottingen. 

\section*{Appendix: The generating function for the harmonic oscillator}
The following path integral is evaluated in \cite{feynman:72}:
\beq
\label{a1}
F[f;x,x']=\int_{x(0)=x}^{x(U)=x'}{\cal D}x(u)
\exp\left(-\int_0^U du\left(\frac{m}{2}\dot{x}^2
+\frac{m\omega^2}{2}x^2+if(u)x(u)
\right)\right)\nonumber\\
\eeq
The r.h.s. of Eq.(\ref{a1}) is given by:
\beq
F[f;x,x']=\sqrt{m\omega/(2\pi\sinh(\omega U))}
e^{-\Phi}
\eeq
where
\beqa
\Phi &=& \frac{1}{4m\omega}\int_0^U du \int_0^U du' e^{-\omega |u-u'|}f(u)f(u')
+\frac{m\omega}{2\sinh(\omega U)}
\left[(x^2+x'^2)\cosh(\omega U) - 2xx'
\right.\nonumber\\
&+&\left.2A(xe^{\omega U}-x')+2B(x'e^{\omega U}-x)
+(A^2+B^2)e^{\omega U}-2AB\right]
\eeqa
where
\beqa
A&=&\frac{i}{2m\omega}\int_0^U du e^{-\omega u}f(u)\nonumber\\
B&=&\frac{i}{2m\omega}\int_0^U du e^{-\omega (U-u)}f(u)
\eeqa

\end{document}